# Photon level crosstalk between parallel fibers installed in urban area

**Mikio Fujiwara,[1,*] Shigehito Miki,[2] Taro Yamashita,[2] Zhen Wang,[2] and Masahide Sasaki[1]**

[1] *National Institute of Information and Communications Technology, 4-2-1 Nukui-Kita, Koganei, Tokyo 184-8795, Japan*

[2] *Kansai Advanced Research Center, National Institute of Information and Communications Technology, 588-2, Iwaoka, Iwaoka-cho, Nishi-ku, Kobe, Hyogo 651-2492, Japan*

[*] *fujiwara@nict.go.jp*

**Abstract:** We estimate the channel characteristics of field-installed dark fibers at single photon levels for quantum key distribution (QKD). Measured fibers are telecom single-mode dark fibers over 45 km connecting a center and suburbs of Tokyo. Their total losses are about 14dB, and 50% of the whole lengths are aerial lines. We find that stray light from other public internet fibers is dominant and crosstalk occurs at bending points in laying cables. These results mean the crosstalk from public networks can increase the bit error rate in the QKD system, and imply an underlying information leakage through an adjacent covered-fiber.

## References and links

1. C. H. Bennett, G. Brassard, "Quantum Cryptography: Public Key Distribution and Coin Tossing," Proceedings of IEEE International Conference on Computers Systems and Signal Processing, Bangalore India, pp 175-179, December (1984).
2. H-K. Lo, X. Ma, and K. Chen, "Decoy state quantum key distribution," Phys. Rev. Lett. **94**, 230504-1-230504-4 (2005).
3. S. Miki, M. Takeda, M. Fujiwara, M. Sasaki, and Z. Wang, "Compactly packaged superconducting nanowire single-photon detector with an optical cavity for multichannel system," Opt. Express **17**, 23557–23564 (2009).
4. S. Miki, T. Yamashita, M. Fujiwara, M. Sasaki, and Z. Wang, "Multi-channel SNSPD system with high detection efficiency at telecommunication wavelength," Opt. Lett. **35**, 2133-2135 (2010)
5. Japan Gigabit Network 2 plus, http://www.jgn.nict.go.jp/jgn2plus/english/index.html.
6. G. Gol'tsman, O. Okunev, G. Chulkova, A. Lipatov, A. Semenov, K. Smirnov, B. Voronov, A. Dzardanov, C. Williams, and R. Sobolewski, " Picosecond superconducting single photon detector," Appl. Phys. Lett. **79**, 705–707 (2001).
7. A. Korneev, P. Kouminov, V. Matvienko, G. Chulkova, K. Smirnov, B. Voronov, G. N. Gol'tsman, M. Currie, W. Lo, K. Wilsher, J. Zhang, W. Słysz, A. Pearlman, A. Verevkin, and R. Sobolewski, "Sensitivity and gigahertz counting performance of NbN superconducting singlephoton detectors," Appl. Phys. Lett. **84**, 5338–5340 (2004).
8. A. Tanaka, M. Fujiwara, S. W. Nam, Y. Nambu, S. Takahashi, W. Maeda, K. Yoshino, S. Miki, B. Baek, Z. Wang, A. Tajima, M. Sasaki, and A. Tomita, "Ultra fast quantum key distribution over a 97 km installed telecom fiber with wavelength division multiplexing clock synchronization," Opt. Express **16,** 11354–11360 (2008).
9. http://documents.exfo.com/specsheets/fcd-10b-ang.pdf
10. http://www.optigate.jp/products/cable/szcable.html
11. D. C. Chang and E.F. Kuester, "Radiation and propagation of a surface-wave mode on a curved open waveguide of arbitrary cross section," Radio Sci. **11**, 449-457(1976).
12. A.R. Dixon, Z. L. Yuan, J. F. Dynes, A. W. Sharpe, and A. J. Shields, "Continuous operation of high bit rate quantum key distribution," Appl. Phys. Lett. **96**, 161102-1–161102-3, 19126 (2010).

## 1. Introduction

Data theft is on the increase and set to rise dramatically in the upcoming years. Despite its reputation for being more secure than standard wiring or airwaves, the truth is that fiber cabling is just as vulnerable to technical hacks. Therefore, technology for information security has become a critical issue for the advanced information and communications network society. Traditionally, data encryption is performed at Layer 3 and above, based on protocols such as Internet Protocol Security (IPSec) and the Secure Socket Layer (SSL) Virtual Private Network (VPN) solution.

On the other hand data protection technology performed at Layer 1 has been developed in recent years. Quantum key distribution (QKD) is a representative example of physical cryptography. QKD allows two users, Alice and Bob, to communicate in absolute security by combing with Vernam's one-time pad. Unconditional security is guaranteed by the fundamental laws of physics. BB84[1] is the most famous protocol of QKD. In this system, signals originated from single photon pulses emitted by Alice are guaranteed to be secure. Therefore, the key devices for QKD are a single photon emitter and a single photon detector at the telecom wavelength. In particular, low dark count rates of detectors are indispensable for extending QKD in the field. In fact, a powerful method to relax the requirement of a single photon emitter has already been known, which is called the decoy state QKD[2] allowing us to extend the distribution length even using an imperfect single photon emitter (weak coherent pulses). In the decoy state QKD system, the dark count rate of a single photon detector limits the transmission distance.

In addition to a detector, we need low loss and "dark" fibers, because QKD is particularly vulnerable to losses and noises. Practical fibers installed in the field are, however, lossy and noisy. They are built by splicing many short-length fibers belonging to different companies at intervals of a few kilometers in urban area. Many spliced junctions cause significant losses. They are installed not only underground but also through aerial cables which are frequently used due to ease of wiring. So the transmission characteristics are sensitive to weather conditions. Those characteristics measurements are indispensable for designing QKD systems. However, characterization of installed fibers for QKD applications has been lacking.

In this paper, we report on the dark count and its origin in field-installed dark fibers at single photon levels for quantum key distribution in urban Tokyo, Japan. We also estimate the crosstalk between parallel fibers using a high sensitive superconducting single photon detector (SSPD)[3,4]. From the channel estimation results, we show the risk of information leakage even through an adjacent covered-fiber.

## 2. Field-installed fiber: Japan Gigabit Network 2 plus (JGN2 plus)

Measured fibers are telecommunication single-mode dark fibers in part of an optical testbed network, called Japan Gigabit Network 2 plus (JGN2 plus)[5]. The National Institute of Information and Communications Technology (NICT) provides JGN2plus optical testbed link that enables demonstrative experiments for network R&D. The dark fibers under our study are laid from the NICT headquarter at Koganei to the center node at Otemachi (very center of Tokyo). The fiber lengths are about 45 km. There are many splicing points resembling actual business situation. The total attenuation amounts to 14dB on average. About 50% of the fibers are aerial lines. Fluctuations in polarizations are typically 3dB per hour. In addition, these fibers in the Koganei-Otemachi link share the same multi-core cable with other commercial fibers alloted for anonymous users. Figure 1 shows a conceptual diagram of the Koganei-Otemachi link in the JGN2plus.

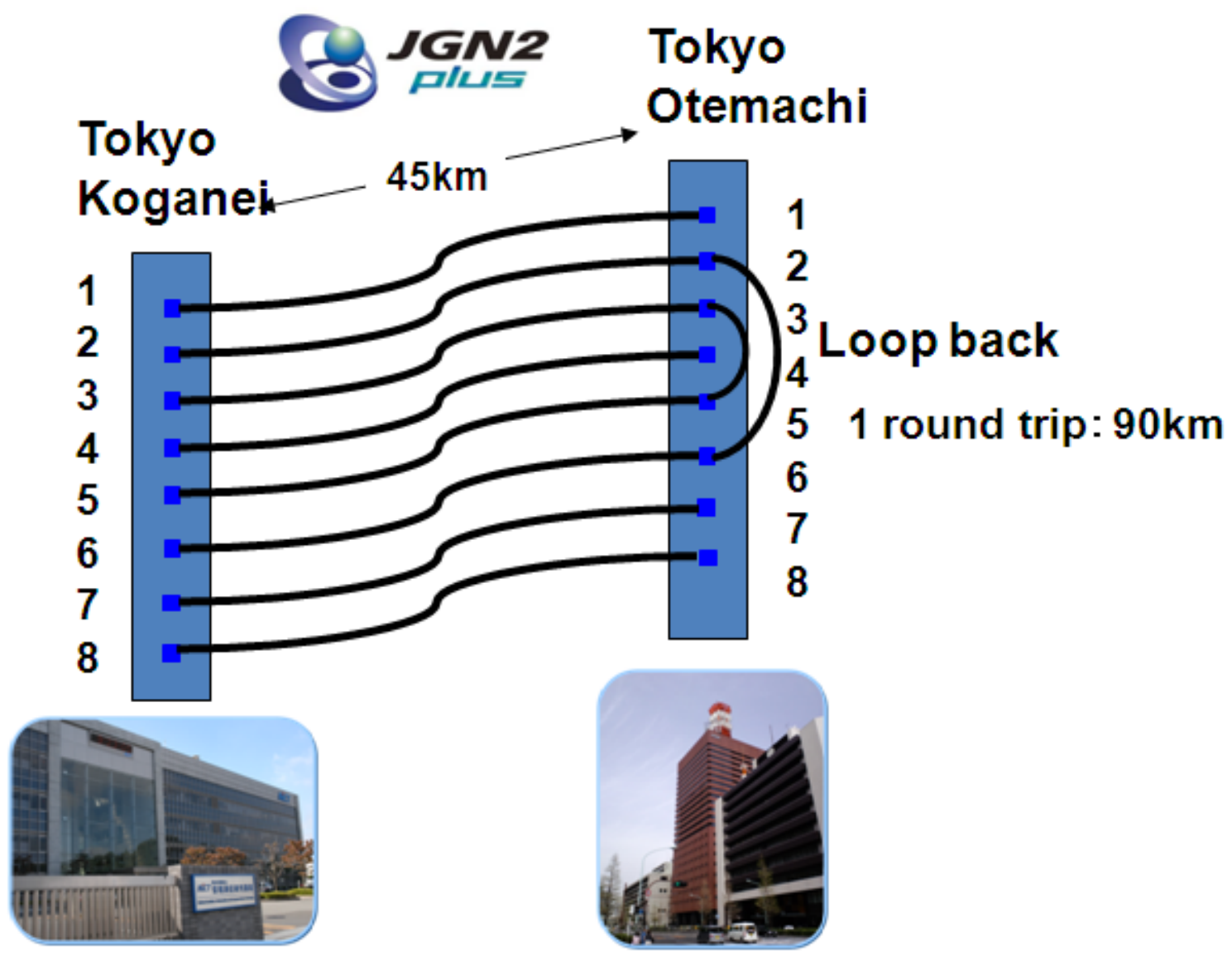

Fig. 1. Conceptual diagram of dark fibers in the Koganei-Otemachi link in the JGN2plus. Eight dark fibers in the cable are allotted to JGN2plus. . The rate of soil is 50%.

## 3. Superconducting single photon detector (SSPD)

We use a NbN SSPD[3,4,6,7] for characterization of dark fibers. We have developed a multi-channel SSPD system based on the Gifford McMahon (GM) cryocooler that can offer guaranteed performance, are cryogen free, and are capable of turnkey, continuous operation with low input power consumption[3,4]. The similar system was used for a QKD experiment through field-installed fibers in 2007[8]. At that time, the single photon detection efficiency (DE) was around 1% at a dark count rate of 100 c/s. This time, however, we use the latest version with enhanced DE by applying an optical cavity structure. Therefore, the DE of the detector has a dependency on the wavelength of input photons. Figure 2 shows the detection efficiency at a dark count of 100 c/s as a function of wavelength. Fringes can be observed and the average detection efficiency is about 15-17%. The SSPD has high sensitivity for the wide wavelength region, including visible light.

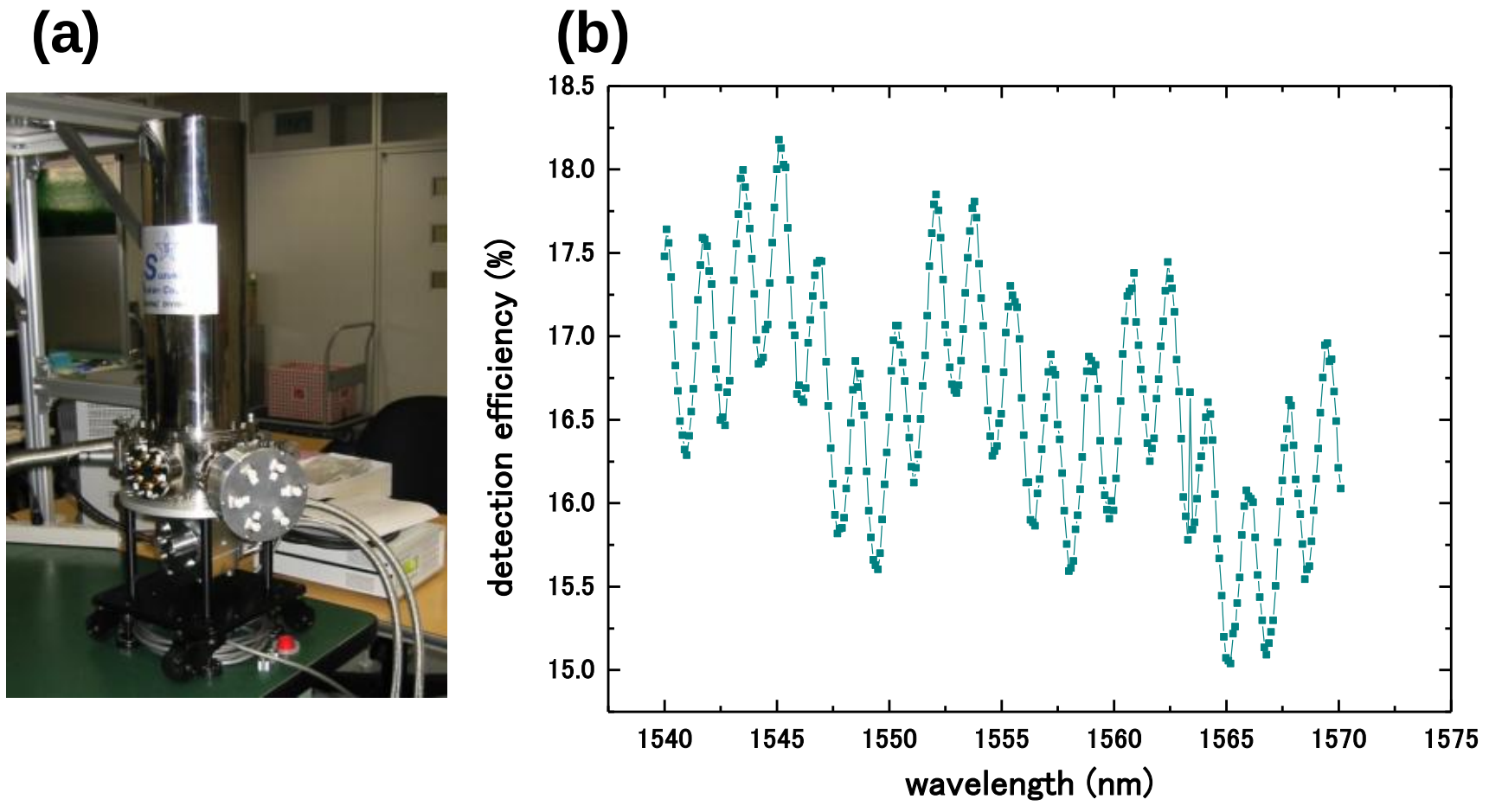

Fig. 2. (a) Picture of a multi-channel SSPD system based on the GM cryocooler. (b) Detection efficiency of the SSPD as a function of wavelength at the dark count rate of 100 c/s.

## 4. Dark count measurement in JGN2plus dark fiber

Experimental setups are depicted in Fig. 3(a)-(c).

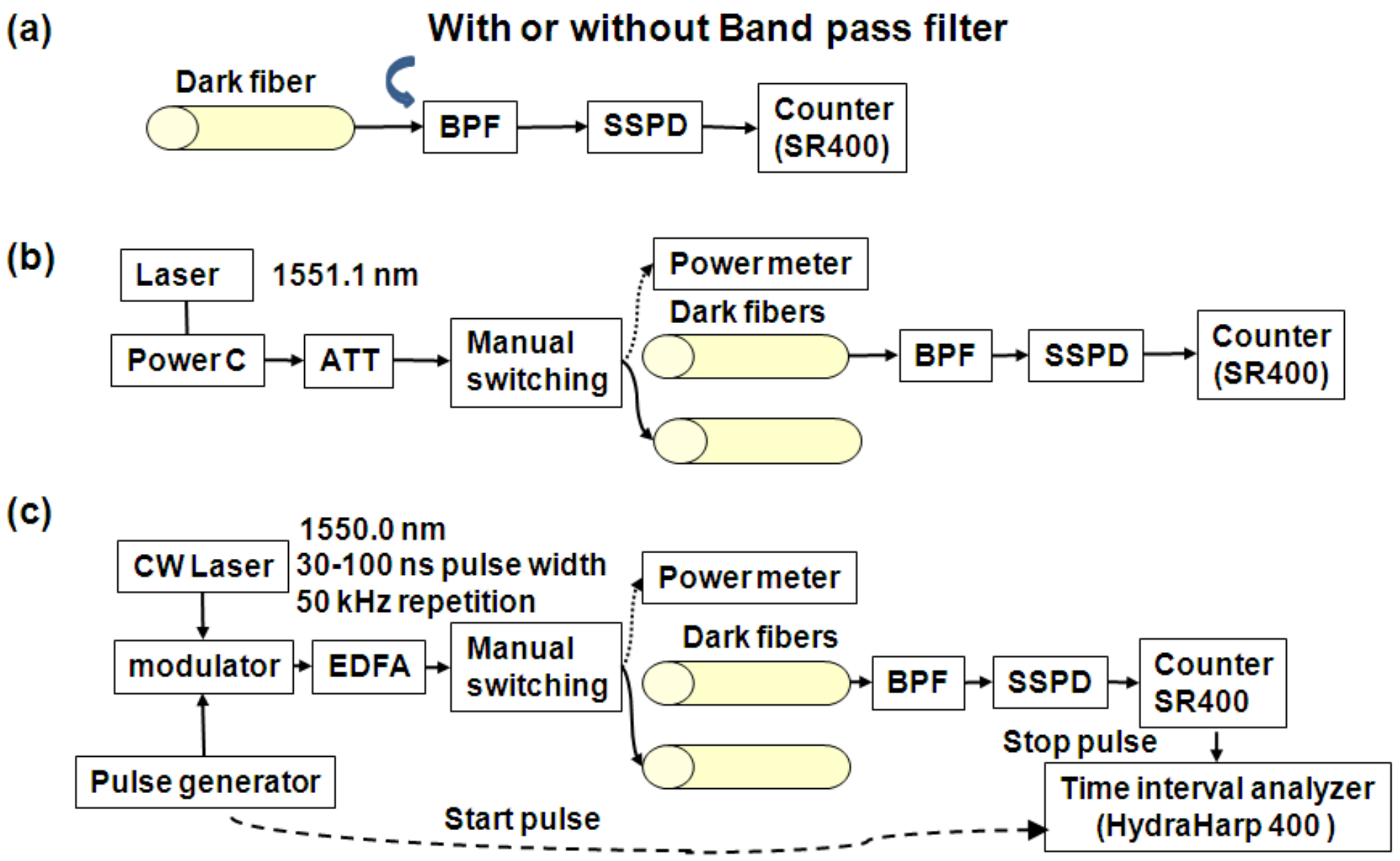

Fig. 3. Conceptual image of fiber characteristic test: (a) Dark count measurement; (b) Crosstalk estimation between neighboring fibers; (3) Identification of leakage points. Power C: power controller, ATT: attenuator, BPF: band pass filter, EDFA: Erbium doped fiber amplifier.

The SSPD is current biased via the dc arm of the bias tee, and the bias current is set at a DE with a 15-17% dark count rate of 100 c/s. The output signals from the ac arm of the bias tee are amplified using a series of two low-noise amplifiers, and then discriminated using a pulse counter (Stanford Research SR400).

In Fig. 4(a), we show dark count rates in daylight and at night in the loopback line over 90 km connecting fiber #2 and #6, when the dark fiber is connected to the SSPD directly. Both are almost the same level around 2000 c/s. This result implies that dark counts due to daylight are not dominant in this fiber, and high dark count rate might be attributed to stray light at telecom wavelengths. To specify the wavelength we measure the spectra using a variable narrow band pass filer as shown in Fig. 3(a). Its attenuation rate and bandwidth are 2dB and 1 nm respectively. Figure 4(b) shows dark count rates as functions of the wavelength in the #2-#6 loopback line, #7, and #8 lines in Fig. 1. A few clear peaks can be recognized in all the fibers around the wavelengths which are widely used in the telecom system. Note that, no light sources are input into the measured fibers and other backup lines. Thus, crosstalk takes place in the cable, and photons in the telecom wavelengths used in other commercial fibers leak into the JGN2plus fibers.

This kind of crosstalk even occurring between covered fibers in a cable should be taken into account in designing QKD systems, even if a dark fiber is dedicated for a quantum channel. Actually, parallel fibers are adopted to realize QKD system in many cases. On the so-called “classical channels,” synchronous signal, base-matching, and/or information for error correction are sent by using blight light signals, which would contaminate the neighboring quantum channel by the crosstalk.

In order to affirm the crosstalk phenomenon between parallel fibers in a quantitative way, we input the CW laser into the #3-#5 loopback line, and measure a dark count rate in the #2-#6 loopback line, which are just neighboring fiber lines, as shown in Fig. 3(c). A significant increase in dark count can be recognized when the input power exceeds -20dBm, as shown Fig. 4(c).

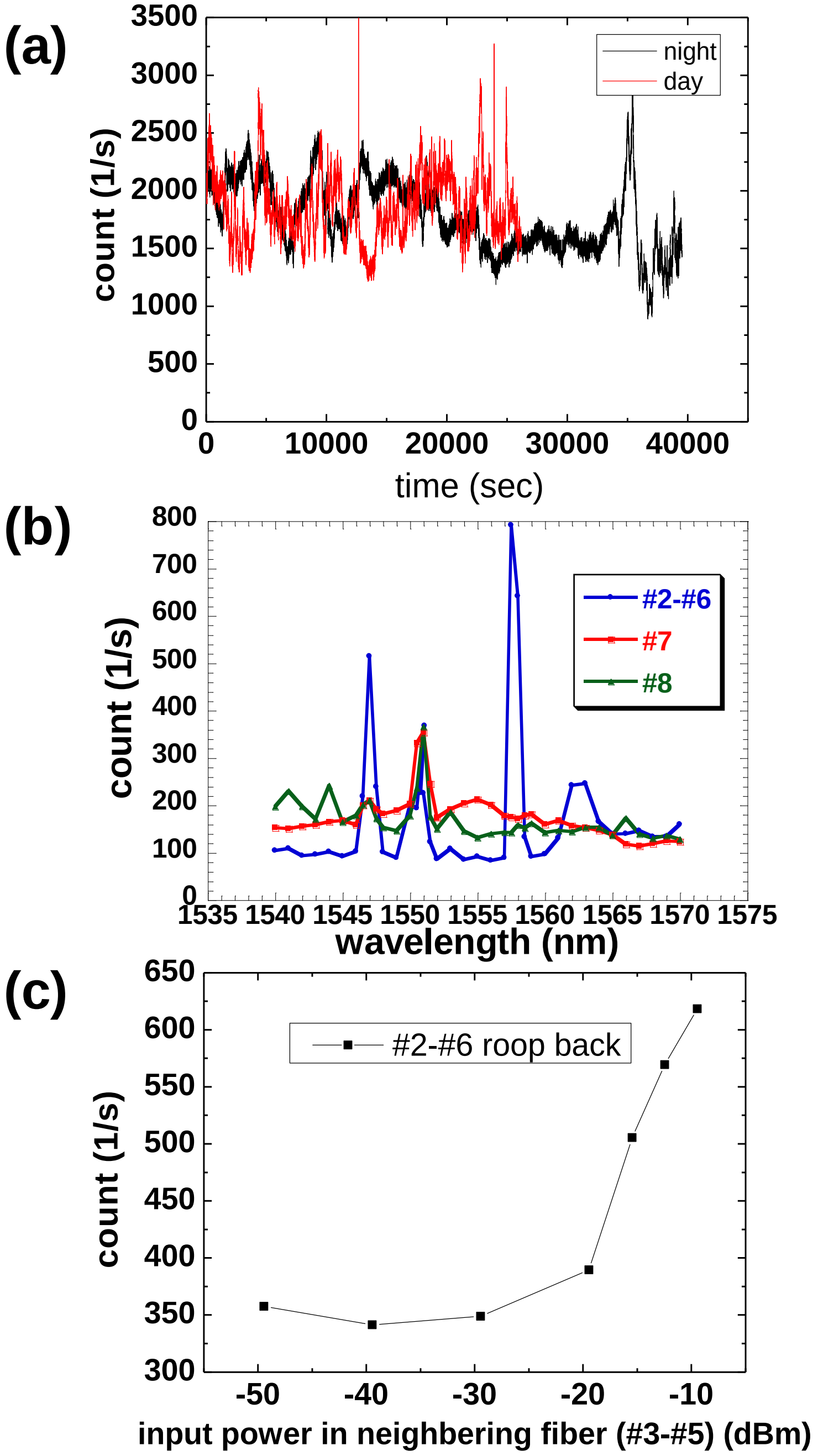

Fig. 4. (a) Dark count rate of JGN2plus fibers without band pass filter; (b) Dark count rate dependency on wavelength; (c) Dark count rate as a function of input power in the neighboring fiber.

## 5. Back scattering of leakage photons

In order to identify where and how the crosstalk between covered fibers takes place, it would be helpful to refer to fiber tapping devices. Actually data theft through a fiber without breaking connection has been reported. The fiber cable to be tapped is stripped and placed into a micro-bend clamping device[9]. The photons leaking from the cable are detected by the photo detector and sent to an optical-electrical converter. Therefore, it is first suspected that photon leakage to the neighboring fiber occurs at a bending part and/or a spliced point that create a large reflection.

So we next try to estimate the leakage points by a method resembling an optical time-domain reflectometer, as shown in Fig. 3(c). Light of 1550 nm from a CW laser is pulse-shaped by a modulator. The pulse width is 30-100 ns and the repetition rate is 50 kHz. This pulse is amplified by an Erbium doped fiber amplifier (EDFA) to 4-7 dBm. When optical pulses are input into the fiber, start pulses are sent to the time-interval analyzer. Stop pulses are given by the SSPD-connected a neighboring fiber when the SSPD detects photons. If the photon leakage occurs at specific points, the peaks of counts at the corresponding back reflecting term should be recognized. Figure 5(a) shows the leakage photon counts distribution as a function of the corresponding transmission length in the #2-#6 loopback line (see Fig. 1) when pulses are input into a neighboring line of the #3-#5 loopback. The pulse width is 100 ns with a 50 kHz repetition rate and the input power is 7 dBm. To compare the fiber characteristics, an optical time-domain reflectrometer (OTDR) trace of the #3-#5 loopback line is shown in Fig. 5(b). The peaks in Fig. 5(a) do not necessarily match those in Fig. 5(b). Namely, splicing points and fiber joints that are recognized in the OTDR trace in Fig. 5(b) due to large reflections, do not match the photon leakage points. Meanwhile, leakage points match cable bending positions (ascertained directly) at least within 1000 m.

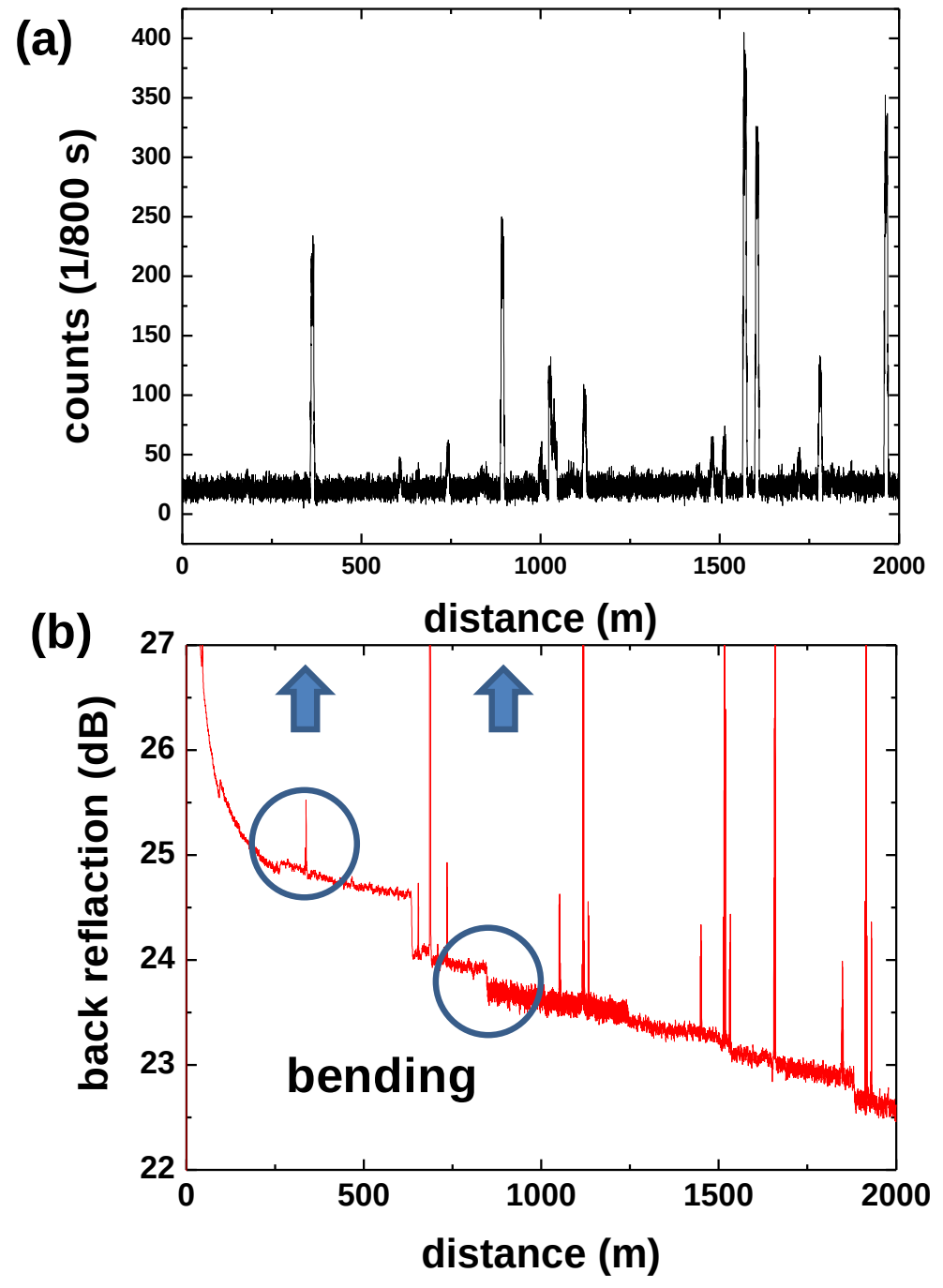

Fig. 5. (a) Back scattering counts of leakage photons as a function of transmission length; (b) OTDR trace of optical pulse input fiber. Large reflection and plunge step in (b) are corresponded to bending positions.

To confirm the bending positions more carefully, we finally measure short-distance fibers in the premise of NICT. The repetition rate of the optical pulse train was 50 kHz when the data in Fig. 5(a) were acquired. Actually, we cannot deny that ghosts could have been measured in Fig. 5(a) under the 50 kHz repetition rate, because it was too fast to estimate the back reflection in a fiber of as long as 90 km. The fiber length in this time is only 700 m, but we can explicitly know the condition of the field-installed fiber directly. Figure 6 shows photon leakage to the nearest neighboring, second neighboring, and third neighboring fiber in the 700 m field-installed fibers. Half a dozen four-core tapes, including the estimated fibers in this experiment, are bundled into one optical cable[10]. The pulse width is set to 30 ns with the same repetition rate of 50 kHz, and the input power is 4 dBm. In the nearest neighboring and second neighboring fibers, reflections of leakage photons are observed at the 130 meter mark. Around this point, the optical cable is bent in a hand-hole, while there are no splicing points or joint connections. In Fig. 6, significant photon leakage can be measured in the second neighboring fiber at the 680 m mark. At this point, leakage photons cannot be observed in the nearest neighboring fiber. The optical cable is also rolled around this point. Photon leakage cannot be observed in the third neighboring fiber. These leakage photons are not to be found in Fig. 5(a). We should add that data in Fig. 5(a) was obtained with the temperature around 12 degrees Celsius and in Fig. 6 with the temperature around 22 degrees Celsius. These results imply that the fiber cable bending is identified as the cause of crosstalk between parallel fibers, and photon leakage reaches at least the second neighboring fiber. Moreover, the leakage points vary depending on weather conditions.

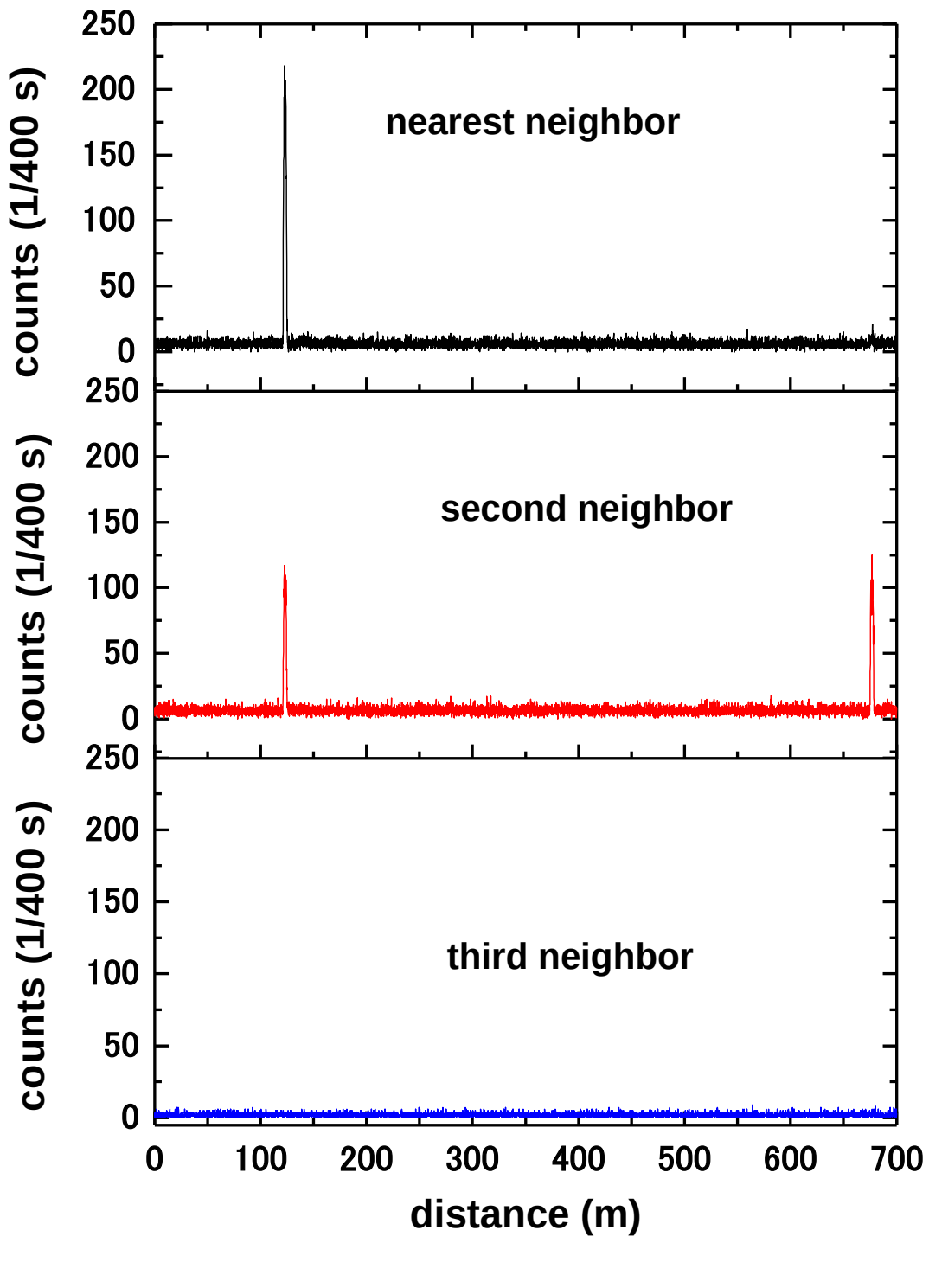

Fig. 6. Extent of leakage photon in parallel fibers

## 6. Discussion

We extrapolate that crosstalk occurs mainly at damping locations, such as splicing or bending points. Based on the results in Fig. 6, we believe that bending of an optical cable is a cause of photon crosstalk between parallel fibers. Bending of an optical cable is a known factor in transmission loss. The loss photons are radiated due to the degradation of the confinement effect. Homogeneous bending loss $\alpha_B$ is given by[11],

$$\alpha_B = \sqrt{\pi}u^2 \exp\left[-4\Delta w^3 R/\left(3a\nu^2\right)\right] / \left[2w^{3/2}V^2(Ra)^{1/2}K_1^2(w)\right] \qquad (1)$$

where $u$ and $w$ are normalization factors of propagation constants, $R$: bending radius, a: fiber core radius, $\Delta$: relative index difference, $V=2\pi n_0 a(2\Delta)^{1/2}/\lambda$ ($n_0$: clad glass index, $\lambda$: wavelength), and $K_1$: first-order modified Bessel function. Therefore, $\alpha_B$ depends on $R$ strongly. The allowable bending radius of our optical cable is 100 mm and the cable is laid with a radius of more than 200 mm. However, crosstalk is observed around the rolled parts. The reason why the crosstalk depends on weather conditions can be explained in such a way that expansion or shrinking of a fiber coating material due to the ambient temperature change photon coupling efficiency. A detailed study of this phenomenon must be conducted to provide a clear explanation about crosstalk probability.

Let us evaluate the influence of stray light in JGN2plus fibers to a QKD system According to the dark count rate shown in Fig. 4(a), if the InGaAs APD with a DE of 15% at 1 ns time-width of Geiger mode is applied to a QKD system, the dark count probability is estimated $2\times10^{-6}$ (= 2000 c/s $\times10^{-9}$ s), even if the APD has no dark count. This value is 20% of a widely used APD dark count probability[12] and large enough to deteriorate a QKD system, because the decoy state QKD is susceptible to dark counts. Therefore, stray light in JGN2plus fibers is a significant barrier to a QKD system.

In addition, a fact that photons leak from a parallel fiber means that the classical channel parallel to the quantum channel sending data of synchronous signal, base-matching, and/or information for error correction would be a potential source of dark count noise. We cannot conclude that this is a universal phenomenon in all the installed fibers, but the potential risk of crosstalk should be taken account and appropriate design consideration must be applied, i.e. tuning input power, wavelength, and method of synchronization. Moreover, fibers should be selected after a thorough assessment of the transmission path.

## 7. Conclusion

We report the crosstalk phenomena between parallel fibers in a cable using a sensitive single photon detector in JGN2plus dark fibers. In field-installed fibers, photons at telecom wavelengths are observed from other commercial networks. Apparently, this crosstalk is discovered only if high sensitive single photon detectors are used. However, the photon leakage is large enough to deteriorate a QKD system. We also point out the need to estimate a channel characteristic, proper selection of wavelengths and narrow band filters when the QKD system is installed in practical network in urban area. Moreover, the probability of photon coupling to the neighboring fiber becomes higher at a bending point. At least the second neighbor in a four-core tape is vulnerable to photon leakage. This is a new risk of information leakage through a parallel fiber. Even without relying upon a fiber tapping device which is currently commercially available[12], and without removing cover materials from fibers, data leakage take place between covered and parallel fibers in an installed cable. In conclusion, data encryption at the physical layer of optical fibers must be indispensable and quantum cryptography as the most trustworthy candidate for it would attract more attention.

**Acknowledgement**

We would like to thank H. Takesue of NTT Basic Research Laboratories and K. Nakamura, K. Terada, K. Negishi, S. Shinada, K. Makino and T. Miyazaki of NICT for their helpful discussion.